\documentclass{article}
\usepackage{spconf,amsmath,amssymb,graphicx,booktabs}
\usepackage{xurl}
\usepackage[table]{xcolor}
\definecolor{oursdark}{RGB}{204,224,247}   
\definecolor{ourslight}{RGB}{233,242,252}  
\usepackage[hidelinks]{hyperref}
\usepackage{orcidlink}

\title{\fontsize{14}{16}\selectfont
GrainSpeech: Less Context, More Detail for Compact Speech Synthesis}
\name{Zitao Liang\orcidlink{0009-0009-1802-4215}, Chang Gao\orcidlink{0000-0002-3284-4078}\sthanks{%
\fontsize{9}{11}\selectfont
Corresponding author: Chang Gao \\
Email: zitaoliang@tudelft.nl, Chang.Gao@tudelft.nl
}}
\address{Department of Microelectronics, Delft University of Technology, The Netherlands}
\begin{document}
%
\maketitle
\begin{abstract}
Compact acoustic models face a challenging quality--capacity trade-off.
We investigate two factors in this regime: encoder context and Mel-spectrogram supervision.
A receptive-field-scaling study shows that expanding self-attention beyond 15 phonemes provides no consistent gains in pitch, energy, or duration prediction. Guided by this finding, we introduce a fixed-receptive-field convolutional encoder that reduces the respective prediction errors by 36.0\%, 17.3\%, and 3.4\%.
We further show that directly transferring image-domain gradient-variance supervision restores fine-scale variation but degrades predicted quality, motivating a Mel-specific formulation with axis-specific gradients, overlapping local statistics, and log-domain
variance matching.
GrainSpeech contains only 264.8K parameters and achieves 17.9$\times$ real-time Mel generation on a microcontroller (MCU), while attaining UTMOS scores comparable to substantially larger models with less than 1.5\% of their parameters. Source code and demos are available at \url{https://github.com/lab-emi/GrainSpeech}.

\end{abstract}


\section{Introduction}
\label{sec:intro}
A common neural text-to-speech (TTS) system uses an acoustic model to
map text to a Mel spectrogram and a neural vocoder to reconstruct the
waveform~\cite{ren2020fastspeech,kong2020hifi}.
However, many high-quality acoustic models contain millions of
parameters, limiting deployment on memory- and computation-constrained
devices~\cite{ren2020fastspeech,tatanov2022mixer,mehta2024matcha}.
EfficientSpeech (ES) demonstrated a compact acoustic model with approximately
266K parameters~\cite{atienza2023efficientspeech}, showing the
feasibility of highly compact acoustic modeling.
Under such tight capacity constraints, maintaining synthesis quality
remains challenging.

We focus on two potential bottlenecks in compact acoustic modeling.
First, the role of encoder context has not been well isolated.
Self-attention encoders commonly provide broad phoneme context
~\cite{ren2020fastspeech,atienza2023efficientspeech},
while convolutional alternatives have also shown strong performance in
acoustic-feature prediction
~\cite{tatanov2022mixer,vainer2020speedyspeech}.
However, prior comparisons typically differ in both encoder architecture
and accessible context, making the contribution of each factor difficult
to isolate.
Second, Mel-spectrogram over-smoothing can degrade synthesized speech
quality~\cite{ren2022revisiting,kogel2023towards}.
Pointwise $L_1$ supervision is prone to this problem, while structural
objectives such as SSIM improve local reconstruction but do not
explicitly supervise fine-grained variation ~\cite{vainer2020speedyspeech,wang2004image}.
Gradient variance (GVar) provides local-variation supervision~\cite{abrahamyan2022gradient},
but its original formulation is designed for spatial image structure.

To address these issues, we propose GrainSpeech, a 264.8K-parameter
acoustic model with a fixed-receptive-field convolutional encoder and
Mel-adapted GVar supervision. Our main contributions are:
\begin{itemize}
\item A  receptive-field study separates
context from architecture: self-attention beyond 15 phonemes gives no
consistent gain, and a parameter-matched fixed-receptive-field
convolutional encoder reduces pitch, energy, and duration errors by
36.0\%, 17.3\%, and 3.4\%.
\item We propose Mel-GVar, which adapts image-domain gradient-variance
supervision to Mel spectrograms via axis-specific gradients, overlapping
local statistics, and log-domain variance matching. Direct transfer
degrades UTMOS to 2.769, whereas Mel-GVar raises it to 4.086.
\item GrainSpeech, having only 264.8K parameters, improves
UTMOS by 0.496 over ES-Tiny at the same budget and is statistically on
par with MixerTTS with 75.7$\times$ fewer parameters, while generating
Mel at 17.9$\times$ real time on an STM32H747XI MCU.
\end{itemize}

\begin{figure*}[t]
  \centering
  \includegraphics[width=\linewidth]{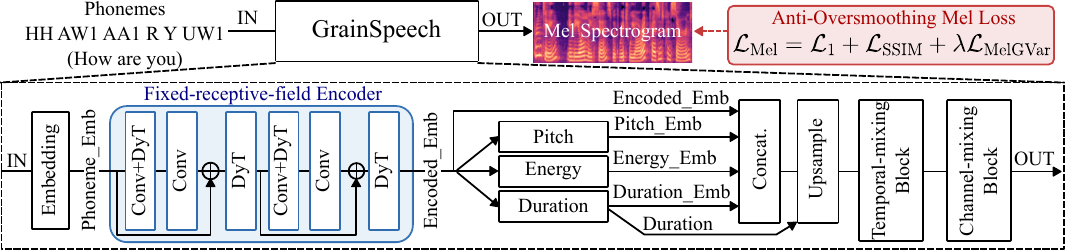}
  \caption{GrainSpeech architecture with a fixed-receptive-field encoder, and an anti-oversmoothing Mel loss.}
  \label{fig:overview}
\end{figure*}

\section{PROPOSED METHOD}

\subsection{System Overview}

Figure~\ref{fig:overview} illustrates GrainSpeech, where a
fixed-receptive-field convolutional encoder predicts pitch, energy, and
duration from the phoneme sequence.
The predicted features are embedded, concatenated with the encoder
output, and upsampled according to duration before being processed by
dilated temporal mixing~\cite{kong2020hifi} and a channel-mixing
bottleneck MLP to generate the Mel spectrogram.
Training uses the anti-oversmoothing Mel objective in
Section~\ref{ssec:gvar}.

\subsection{Fixed-Receptive-Field Encoder}

To examine the role of encoder receptive field in acoustic-feature
prediction, we construct a single-block self-attention encoder based on
EfficientSpeech~\cite{atienza2023efficientspeech}.
We define $W$ as the encoder receptive field on the original phoneme
sequence and control it using symmetric attention masks.
Here, $W$ refers only to the encoder; downstream acoustic-feature
predictors further expand the context equally across all
variants.

Motivated by this context study, we construct a fixed-receptive-field
convolutional encoder for matched-context comparison and the final
GrainSpeech model. As shown in Fig.~\ref{fig:overview}, it consists of
a phoneme embedding layer followed by stacked residual convolutional
blocks, with dilation schedules chosen to realize different $W$ values.
This allows attention and convolutional encoders to be compared under
the same accessible phoneme context. Following prior work, LayerNorm is
replaced with DynamicTanh (DyT) in the convolutional blocks
~\cite{zhu2025transformers,slimtts}.

\subsection{Anti-Oversmoothing Mel Loss}
\label{ssec:gvar}
Mel-spectrogram over-smoothing suppresses fine-grained acoustic
variation and can degrade synthesized speech quality
~\cite{ren2022revisiting}. 
To explicitly supervise these local variations, we adapt the gradient-variance (GVar) objective originally proposed for image super-resolution~\cite{abrahamyan2022gradient}. 
The original GVar uses Sobel gradients, non-overlapping spatial
patches, and raw-domain $L_2$ variance matching.
Since Mel spectrograms represent time and Mel frequency rather than two
spatial dimensions, we retain the local gradient-variance principle
while adapting three components:
(1) axis-wise first-order differences replace Sobel filtering to
characterize temporal and spectral variations;
(2) sliding $11\times5$ neighborhoods replace non-overlapping patches
to provide overlapping local statistics; and
(3) log-domain $L_1$ matching replaces raw-domain $L_2$ matching to
compare relative rather than absolute variance discrepancies.

Let $\widehat{\mathbf{M}},\mathbf{M}\in\mathbb{R}^{T\times F}$ denote
the predicted and reference Mel spectrograms. We define
\begin{equation}
\begin{aligned}
G_t(\mathbf{M})_{t,f}&=\mathbf{M}_{t+1,f}-\mathbf{M}_{t,f},\\
G_f(\mathbf{M})_{t,f}&=\mathbf{M}_{t,f+1}-\mathbf{M}_{t,f}.
\end{aligned}
\end{equation}

Let
$S_d(\mathbf{M})=
\log(\operatorname{LVar}_{11\times5}(G_d(\mathbf{M}))+10^{-6})$,
where $\operatorname{LVar}_{11\times5}$ is population variance over
unit-stride time$\times$Mel-frequency windows, excluding padded elements.
We define
\begin{equation}
\mathcal{L}_{\mathrm{MelGVar}}
=
\frac{1}{2}\sum_{d\in\{t,f\}}
\|S_d(\widehat{\mathbf{M}})-S_d(\mathbf{M})\|_{1,\Omega},
\end{equation}
where $\Omega$ denotes valid non-padded positions.

The Mel objective is
$\mathcal{L}_{\mathrm{Mel}}
=\mathcal{L}_1+\mathcal{L}_{\mathrm{SSIM}}
+\lambda\mathcal{L}_{\mathrm{MelGVar}}$.

\begin{figure*}[t]
    \centering
    \includegraphics[width=0.98\linewidth]{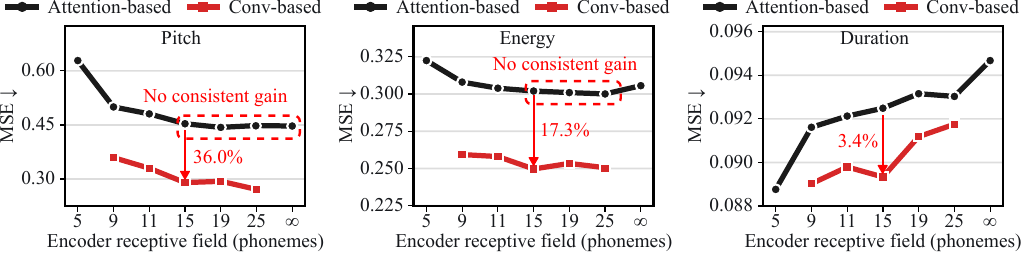}
    \caption{Effects of encoder receptive field and architecture on
    acoustic-feature prediction. MSE denotes mean squared error.}
    \label{fig:window}
\end{figure*}

\section{Experiments}

\subsection{Experimental Setup}

We use LJSpeech~\cite{ljspeech17}, with 12,588 utterances for training
and the remaining 512 split into 384 validation and 128 evaluation
utterances. All variants use the same training seed and are evaluated
after 5,000 epochs.
Phonemes and durations are obtained from forced-alignment
TextGrids~\cite{mcauliffe2017montreal}. WORLD~\cite{morise2016world} pitch and energy are
averaged by phoneme and normalized over the training set.
Audio is sampled at 22.05~kHz and converted to 80-bin Mel spectrograms
using a 1,024-point FFT and window, a hop length of 256, and a
0--8~kHz frequency range.
Our models are trained from scratch using AdamW with batch size 128,
learning rate $10^{-3}$, weight decay $10^{-5}$, 50 warm-up epochs,
and cosine decay.

All synthesized Mels are vocoded using the same
HiFi-GAN-v2~\cite{kong2020hifi}. Speech quality is evaluated using
UTMOS~\cite{saeki2022utmos}. Intelligibility is measured by WER using
Whisper-large-v3~\cite{radford2023robust} with fixed greedy decoding and identical English text normalization for all systems. 
Spectral distortion is measured by MCD-DTW~\cite{kubichek1993mel} using 24-dimensional mel-cepstral coefficients ($c_1$--$c_{24}$, $\alpha=0.455$) extracted from WORLD spectral envelopes at a 5-ms frame shift and aligned by dynamic time warping. 
WER and MCD-DTW are evaluated on the same 128 utterances, with
95\% confidence intervals obtained from 10,000 shared-index
bootstrap replicates.

\subsection{Encoder Receptive Field and Architecture Study}

To separate the effects of encoder context and architecture, we compare
parameter-matched attention and convolutional encoders while keeping all
downstream components and the Mel objective unchanged.
For attention, we evaluate
$W\in\{5,9,11,15,19,25,\infty\}$, where $W=\infty$ denotes global 
context; for convolution, we evaluate
$W\in\{9,11,15,19,25\}$.
The two encoders contain 147,368 and 147,428 parameters, respectively.
$W=15$ is selected based on validation acoustic-feature errors and is
used for the final GrainSpeech. 

As shown in Fig.~\ref{fig:window}, expanding the attention receptive
field beyond $W=15$ provides no consistent improvement in pitch,
energy, or duration prediction, indicating limited benefit from
additional encoder context under this setting.
At matched receptive fields, the convolutional encoder achieves lower
MSE across all three tasks and all shared $W$ values.
At $W=15$, it reduces pitch, energy, and duration MSE by 36.0\%,
17.3\%, and 3.4\%, respectively.
These results show that the convolutional encoder provides more
effective acoustic-feature prediction under a matched parameter budget.

\begin{figure*}[t]
    \centering
    \includegraphics{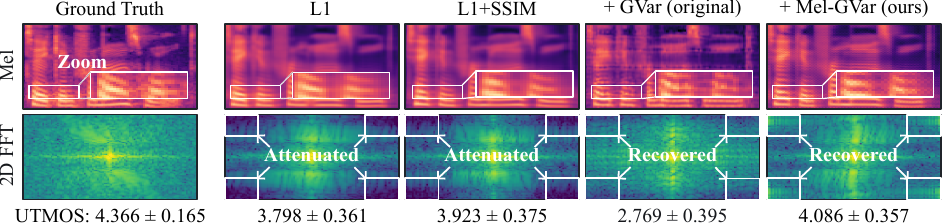}
    \caption{Mel spectrograms and normalized 2D Fourier spectra for GrainSpeech under different Mel losses. Values are mean ± std over 128 evaluation utterances.}
    \label{fig:2dfft}
\end{figure*}

\subsection{Oversmoothing Analysis and Loss Ablation}
\label{ssec:loss_ablation}

We use the SSIM formulation and settings of
~\cite{vainer2020speedyspeech}.
For the original-GVar baseline, we use the original Sobel-gradient,
non-overlapping-patch, and raw-domain $L_2$ formulation
~\cite{abrahamyan2022gradient}, with
$\lambda_{\mathrm{GVar}}=0.01$ following the official implementation.
Our Mel-GVar uses $\lambda_{\mathrm{GVar}}=0.5$.

For visualization, we mean-center each Mel spectrogram, apply a 2D
Hann window, and compute its normalized 2D Fourier power spectrum.
Figure~\ref{fig:2dfft} compares the same utterance under four Mel
objectives. Relative to the ground truth, $L_1$ predictions exhibit
smoother Mel patterns and attenuated energy in the corners of the 2D
spectrum. Adding SSIM improves UTMOS from 3.798 to 3.923, but the
corner-region attenuation remains visible.

Directly applying the original image-domain GVar restores substantial
corner-region spectral energy and fine-scale variation, but introduces
visible horizontal discontinuities that perturb the high-energy
structures of the Mel spectrogram and reduces UTMOS to 2.769.
This contrast suggests that recovering fine-scale variation alone is
insufficient; the recovered variation should also preserve coherent
time--frequency structure.
In comparison, the proposed Mel-GVar restores fine-scale variation
while maintaining more continuous Mel structures and increases UTMOS
to 4.086, the highest among the evaluated objectives.
These results support adapting the image-domain GVar formulation to
Mel spectrograms.

To examine whether the proposed Mel objective is specific to the
GrainSpeech architecture, we apply it to ES-Tiny.
It improves ES-Tiny UTMOS from 3.591 to 3.945 and MCD-DTW from
6.374 to 6.350, while WER remains similar (3.08\% vs.\ 3.13\%).
These gains suggest that its benefit is not specific to the proposed encoder.

\begin{table*}[t]
\centering
\caption{Model size and objective quality on 128 evaluation utterances; brackets are 95\% CIs from 10,000 utterance-level bootstrap replicates. ``Size vs.\ ours'' is the parameter count relative to GrainSpeech.
{\color{oursdark}\rule{1.4ex}{1.4ex}}~GrainSpeech (this work);
{\color{ourslight}\rule{1.4ex}{1.4ex}}~ES-Tiny retrained with our Mel-GVar loss only.
$^\dagger$UTMOS CI overlaps with GrainSpeech at 75.7$\times$ the parameters.}
\label{tab:quality}
\vspace{1mm}
\fontsize{9}{10}\selectfont
\setlength{\tabcolsep}{7.0pt}
\renewcommand{\arraystretch}{1.05}

\begin{tabular}{lrrrrr}
\toprule
Model
& Params. $\downarrow$
& Size vs.\ ours
& UTMOS $\uparrow$
& WER (\%) $\downarrow$
& MCD-DTW (dB) $\downarrow$ \\
\midrule

\rowcolor{oursdark}
\textbf{GrainSpeech (this work)}
& \textbf{0.265M}
& \textbf{1$\times$}
& 4.086 [4.020, 4.148]
& 3.27 [2.32, 4.32]
& 6.314 [6.255, 6.374] \\

\rowcolor{ourslight}
ES-Tiny~\cite{atienza2023efficientspeech} + our Mel-GVar loss
& 0.266M
& 1.0$\times$
& 3.945 [3.872, 4.016]
& 3.13 [2.11, 4.31]
& 6.350 [6.285, 6.419] \\

ES-Tiny~\cite{atienza2023efficientspeech}
& 0.266M
& 1.0$\times$
& 3.591 [3.522, 3.658]
& 3.08 [2.08, 4.23]
& 6.374 [6.314, 6.439] \\

ES-Small~\cite{atienza2023efficientspeech}
& 0.952M
& 3.6$\times$
& 3.885 [3.820, 3.949]
& 3.13 [2.23, 4.15]
& 6.286 [6.231, 6.344] \\

ES-Base~\cite{atienza2023efficientspeech}
& 3.953M
& 14.9$\times$
& 3.900 [3.836, 3.963]
& 3.50 [2.53, 4.57]
& 6.183 [6.120, 6.250] \\

SpeedySpeech~\cite{vainer2020speedyspeech}
& 4.306M
& 16.2$\times$
& 3.716 [3.663, 3.769]
& 4.19 [3.18, 5.29]
& 6.352 [6.285, 6.421] \\

MatchaTTS~\cite{mehta2024matcha}
& 18.204M
& 68.7$\times$
& \textbf{4.264 [4.225, 4.300]}
& 1.98 [1.31, 2.73]
& 5.597 [5.548, 5.653] \\

MixerTTS~\cite{tatanov2022mixer}
& 20.060M
& 75.7$\times$
& 4.087$^\dagger$ [4.030, 4.142]
& \textbf{1.80 [1.12, 2.59]}
& \textbf{5.374 [5.327, 5.426]} \\

FastSpeech~2~\cite{ren2020fastspeech}
& 35.159M
& 132.7$\times$
& 3.953 [3.889, 4.015]
& 2.85 [1.95, 3.84]
& 5.935 [5.867, 6.008] \\

\midrule
Ground Truth
& --
& --
& 4.366 [4.340, 4.389]
& 1.98 [1.28, 2.77]
& -- \\

\bottomrule
\end{tabular}
\end{table*}

\begin{table}[t]
\centering
\caption{A16W8 MCU deployment on an STM32H747XI.
A16W8 denotes 16-bit activations and 8-bit weights; mRTF is generated
duration divided by acoustic-model inference time; $\Delta$UTMOS is the
change from FP32; SRAM is peak usage.}
\label{tab:mcu}
\vspace{1mm}
\fontsize{9}{10}\selectfont
\setlength{\tabcolsep}{3.5pt}
\renewcommand{\arraystretch}{0.95}

\begin{tabular}{lccc}
\toprule
Model & mRTF $\uparrow$ & $\Delta$UTMOS $\uparrow$ & SRAM $\downarrow$ \\
\midrule
\rowcolor{oursdark}
\textbf{GrainSpeech (this work)}
& \textbf{17.9} & $-0.036$ & 483.19 KiB \\
ES-Tiny~\cite{atienza2023efficientspeech}
& 17.2 & $-0.079$ & \textbf{478.59 KiB} \\
ES-Small~\cite{atienza2023efficientspeech}
& -- & $\mathbf{-0.019}$ & Out-of-Mem. \\
\bottomrule
\end{tabular}
\end{table}

\subsection{Overall Comparison}
\label{ssec:overall}

Table~\ref{tab:quality} compares model size and objective quality on the
128 evaluation utterances with a common vocoder and evaluation pipeline.
At the ES-Tiny parameter budget, GrainSpeech improves UTMOS by 0.496
(paired-bootstrap 95\% CI: [0.431, 0.562]) with a close WER
(3.27\% vs.\ 3.08\%) and slightly lower MCD-DTW (6.314 vs.\ 6.374).
The two shaded rows separate the sources of this gain: our Mel-GVar
loss alone lifts ES-Tiny from 3.591 to 3.945 UTMOS, and the encoder adds
a further 0.141 under the same objective (paired-bootstrap 95\% CI:
[0.077, 0.204]).

GrainSpeech attains the highest UTMOS below 5M parameters and is
statistically indistinguishable from MixerTTS (4.086 vs.\ 4.087,
overlapping CIs) with 75.7$\times$ fewer parameters; only MatchaTTS
scores higher, at 68.7$\times$ the size. The 18--35M-parameter models
still obtain lower WER and MCD-DTW.

Table~\ref{tab:mcu} reports A16W8 deployment on an STM32H747XI
(480\,MHz Cortex-M7, 1\,MB SRAM).
GrainSpeech generates Mel spectrograms at 17.9$\times$ real time within
483.19\,KiB peak SRAM, matching the ES-Tiny footprint within 1\% at
slightly higher throughput, and loses only 0.036 UTMOS to quantization
versus 0.079 for ES-Tiny.
ES-Small already exceeds the available SRAM, so the quality gain is
obtained within the largest ES footprint that fits this MCU.

\section{Discussion}
\label{sec:discussion}

Two caveats apply to Table~\ref{tab:quality}: evaluation relies on
automatic metrics on LJSpeech, so listening tests and broader datasets
are needed to confirm the UTMOS gains; and the external checkpoints use
their native data splits, training settings, and frontends, so their
rows are reference points rather than controlled comparisons, whereas
the ES-Tiny rows share our pipeline.
The context study is specific to the evaluated architecture and
training setting and does not establish a universal receptive-field
requirement.
Mel-GVar constrains local variation statistics rather than exact detail
locations, so recovered details need not match the reference point by
point; its three adaptations are evaluated jointly, and their
individual contributions remain to be isolated.
Finally, Table~\ref{tab:mcu} covers Mel generation only; end-to-end
deployment including vocoding is beyond our scope.

\section{Conclusion}
\label{sec:conclusion}
We presented GrainSpeech, a 264.8K-parameter acoustic model. Under our experimental setting, additional self-attention encoder context beyond
15 phonemes provides no consistent benefit, while the matched-context
convolutional encoder improves acoustic-feature prediction.
Our Mel-GVar restores fine-grained variation while avoiding the quality degradation observed with direct transfer of the image-domain objective.
At the ES-Tiny budget, GrainSpeech raises UTMOS from 3.591 to 4.086,
statistically comparable to MixerTTS with 75.7$\times$ fewer parameters,
and generates Mel at 17.9$\times$ real time within 483\,KiB of SRAM on
an STM32H747XI MCU.

\vfill\pagebreak

\section{Compliance with Ethical Standards}
This study uses the publicly available LJSpeech dataset~\cite{ljspeech17}. No participants were recruited,
no new recordings were collected, and no human listening tests or animal experiments were conducted.

\section{Acknowledgments}
This work was partially supported by the Dutch Research Council (NWO) under the Talent Programme Veni 2023 scheme in Applied and Engineering Sciences (AES), Grant No. 21132 (Energy-Efficient Real-Time Edge Intelligence for Wearable Healthcare Devices).
The authors declare no conflicts of interest.

\bibliographystyle{IEEEbib}
\bibliography{refs}

\end{document}